\documentclass[%
 reprint,
superscriptaddress,
 amsmath,amssymb,
 aps,
]{revtex4-2}

\usepackage{graphicx}% Include figure files
\usepackage{dcolumn}% Align table columns on decimal point
\usepackage{bm}% bold math
\usepackage{hyperref}% add hypertext capabilities
\usepackage[utf8]{inputenc}
\usepackage{fontenc}

\hypersetup{colorlinks=true, linkcolor=blue, citecolor=blue, filecolor=blue, urlcolor=blue}
\begin{document}

%\preprint{APS/123-QED}

\title{Topological nontrivial berry phase in altermagnet CrSb}% Force line breaks with \\


\author{Jianhua Du\textsuperscript{¶}}
%\thanks {These authors contributed equally to this work.}
\email[Contact author:]{jhdu@cjlu.edu.cn}

 \affiliation{Department of Physics, China Jiliang University, Hangzhou 310018, China}
\author{Xin Peng\textsuperscript{¶}}%\thanks{These authors contributed equally to this work.}
\affiliation{Department of Physics, China Jiliang University, Hangzhou 310018, China}
\author{Yuzhi Wang\textsuperscript{¶}}%\thanks{These authors contributed equally to this work.}
\affiliation{Beijing National Laboratory for Condensed Matter Physics and Institute of physics, Chinese Academy of Sciences, Beijing 100190, China}
\affiliation{University of Chinese Academy of Sciences, Beijing 100049, China}

\author{Shengnan Zhang}
\affiliation{Beijing Polytechnic College, Beijing 100042, China}
\author{Yuran Sun}
\affiliation{Department of Physics, China Jiliang University, Hangzhou 310018, China}
\author{Chunxiang Wu}
\affiliation{School of Physics, Zhejiang University, Hangzhou 310027, China}
\author{Tingyu Zhou}
\affiliation{School of Physics, Zhejiang University, Hangzhou 310027, China}
\author{Le Liu}
\affiliation{School of Physics, Zhejiang University, Hangzhou 310027, China}
\author{Hangdong Wang}
\affiliation{School of Physics, Hangzhou Normal University, Hangzhou 311121, China}
\author{Jinhu Yang}
\affiliation{School of Physics, Hangzhou Normal University, Hangzhou 311121, China}
\author{Bin Chen}
\affiliation{School of Physics, Hangzhou Normal University, Hangzhou 311121, China}
\author{Chuanying Xi}
\affiliation{Anhui Province Key Laboratory of Condensed Matter Physics at Extreme Conditions, High Magnetic Field Laboratory, Chinese Academy of Sciences, Hefei 230031, China}
\author{Zhiwei Jiao}\email [Contact author:]{jiaozw@cjlu.edu.cn}
\affiliation{Department of Physics, China Jiliang University, Hangzhou 310018, China}
\author{Quansheng Wu}\email [Contact author:]{quansheng.wu@iphy.ac.cn}
\affiliation{Beijing National Laboratory for Condensed Matter Physics and Institute of physics, Chinese Academy of Sciences, Beijing 100190, China}
\affiliation{University of Chinese Academy of Sciences, Beijing 100049, China}
\author{Minghu Fang}\email [Contact author:]{mhfang@zju.edu.cn}
\affiliation{School of Physics, Zhejiang University, Hangzhou 310027, China}
\affiliation{Collaborative Innovation Center of Advanced Microstructure, Nanjing University, Nanjing 210093, China}

%\footnote[¶] {These authors contributed equally to this work.}

%\author{Ann Author}% 
%\thanks{contribut equally}%
%\altaffiliation{
% 	Physics Department, XYZ University.}%Lines break automatically or can be forced with \\
%\affiliation{%

%\affiliation{
%\affiliation{
%\affiliation{%
% &This line break forced with \textbackslash\textbackslash
%5}%


\date{\today}% It is always \today, today,

\begin{abstract}
The study of topological properties in magnetic materials has long been one of the forefront research areas in condensed matter physics. CrSb, as a prototypical candidate material for altermagnetism, has attracted significant attention due to its unique magnetic properties. This system provides a novel platform for exploring the intrinsic relationship between altermagnetic order and exotic topological states. In this study, we combine systematic electrical transport experiments with first-principles calculations to investigate the possible realization mechanisms of topological semimetal states in CrSb and their manifestations in quantum transport phenomena. Our high-field magnetotransport measurements reveal that the magnetoresistance of CrSb exhibits no sign of saturation up to 35 T, following a distinct power-law dependence with an exponent of 1.48. The nonlinear Hall resistivity further indicates a multiband charge transport mechanism. Under high magnetic fields, we observe pronounced Shubnikov-de Haas (SdH) quantum oscillations and discernible Zeeman-effect-induced band splitting at 1.6 K. Systematic Fermi surface and band calculations combined with Berry phase analysis confirm the nontrivial topological character of this material (with a Berry phase approaching $\pi$). These findings not only provide crucial experimental evidence for understanding the electronic structure of CrSb, but also establish an important foundation for investigating topological quantum states in altermagnets.          
\end{abstract}

                
\maketitle

%\tableofcontents

\section{\label{sec:level1} INTRODUCTION} %\textbackslash\textbackslash}

Altermagnets have recently emerged as a novel class of magnetic materials attracting intense interest due to their unique electronic structures and potential topological properties. Unlike conventional ferromagnets and antiferromagnets, altermagnets maintain zero net magnetization while exhibiting momentum-dependent spin-split bands driven by exchange interactions, a characteristic enabling significant band splitting even without spin-orbit coupling (SOC)\cite{PhysRevX.12.040501,PhysRevX.12.031042, Nature.626.517}. Theoretical studies suggest that this peculiar symmetry breaking may induce nontrivial topological electronic states, such as Weyl or Dirac semimetal phases\cite{jungwirth2024altermagnetsbeyondnodalmagneticallyordered,PhysRevB.109.L201109, PhysRevLett.134.096703,PhysRevB.109.024404,PhysRevLett.133.256601,https://doi.org/10.1002/adfm.202409327}. Weyl semimetals, in particular, have become a research focus owing to their topologically protected surface states and exotic transport properties\cite{doi:10.1126/science.aaa9297,PhysRevB.83.205101,PhysRevLett.107.127205,NatPhys.11.645,doi:10.1126/science.aav2334,PhysRevX.5.031023}. Although most known Weyl materials rely on strong SOC or specific crystal symmetries\cite{RevModPhys.90.015001,doi:10.1126/sciadv.1600295,PNAS.114.10596}, altermagnets, characterized by their distinctive magnetic symmetries and structures, offer promising opportunities for novel coupling with topological states, thereby providing an alternative and potentially versatile platform for realizing topological quantum phases. Thus, investigating topological phenomena in altermagnets is of fundamental significance while enabling novel spintronic applications\cite{https://doi.org/10.1002/adfm.202409327,vila2024orbitalspinlockingopticalsignatures,fu2025allelectrically}.The CrSb compound, crystallizing in a hexagonal NiAs-type structure (space group $P6_{3}/mmc$), serves as a prototypical $g$-wave altermagnet with a high Néel temperature of 712 K\cite{PhysRevB.111.144402}, and pronounced spin-split bands in momentum space\cite{https://doi.org/10.1002/advs.202406529,zhou2024crystaldesignaltermagnetism,PhysRevLett.133.206401}. Although studies using angle-resolved photoemission spectroscopy (ARPES) have observed spin-polarized topological Fermi arcs near the Fermi level, suggesting that CrSb may host a Weyl semimetal state\cite{doi:10.1021/acs.nanolett.5c00482,li2024topologicalweylaltermagnetismcrsb}, direct experimental evidence for its topological properties in charge transport measurements remains limited. Understanding how the altermagnetic order modulates CrSb's electronic structure and influences its topological transport behavior is crucial to unraveling the coupling mechanism between altermagnetism and topology.

In this paper, we successfully grown high-quality CrSb single crystals and systematically investigated their topological quantum states through extreme low-temperature, high-field magneto-transport measurements combined with first-principles calculations. Under magnetic fields up to 35 T, we observe pronounced Shubnikov-de Haas (SdH) oscillations, providing direct evidence of topological transport in this multiband system. The intrinsic spin polarization of altermagnetic order cooperates with external magnetic fields to induce Zeeman splitting in CrSb. Our findings not only confirm the existence of a robust topological semimetal state in CrSb but also establish a material platform for designing next-generation spin-topotronic devices.  

\section{\label{sec:level2}Experimental details}

High-quality bulk single crystals of CrSb were grown using the chemical vapor transport method. High-purity chromium and antimony powders (Alfa Aesar, 99.99\%) were weighed in a 1 : 1 molar ratio, with a total mass of 1 g. The powders were finely ground in an agate mortar and loaded into a quartz tube along with iodine (I$_{2}$, 10 mg/cm$^{2}$) as the transport agent. The tube was evacuated, sealed under vacuum, and placed in a two-zone tube furnace. The source zone was maintained at 950$^{\circ}$C and the growth zone at 850$^{\circ}$C for two weeks. After natural cooling to room temperature, Shiny black single crystals with typical dimensions of 2 × 2 × 0.05 mm$^{3}$ were collected at the cooler end of the tube.The chemical composition of the crystals was confirmed to be close to the stoichiometric (Cr : Sb = 1 : 1) using energy-dispersive x-ray spectroscopy (EDX). Structural characterization was performed using both powder and single-crystal x-ray diffraction (XRD), confirming the expected lattice structure. For transport measurements, electrical resistivity, magnetoresistance and Hall resistivity were measured using the standard four-probe method in a Quantum Design physical property measurement system (PPMS-7T) and a water-cooled magnet with the highest magnetic field up to 35 T. To eliminate spurious signals caused by voltage misalignment, magnetic field polarity was reversed during the measurements.

First-principles density-functional-theory (DFT) calculations were carried out with the Vienna ab initio Simulation Package (VASP)\cite{kresseEfficientIterativeSchemes1996a,kresseUltrasoftPseudopotentialsProjector1999a}, employing the Perdew-Burke-Ernzerhof (PBE) generalized-gradient approximation for the exchange-correlation functional\cite{perdewGeneralizedGradientApproximation1996b}. A kinetic-energy cutoff of 400 eV and an 11 × 11 × 9 k-point mesh were used for the bulk calculations. A Wannier tight-binding model (WTBM)\cite{souzaMaximallyLocalizedWannier2001b, marzariMaximallyLocalizedGeneralized1997b, marzariMaximallyLocalizedWannier2012a} constructed with WANNIER90\cite{mostofiUpdatedVersionWannier902014} was processed in WANNIERTOOLS\cite{wuWannierToolsOpensourceSoftware2018a} to obtain the Fermi surface. The extremal cross-sectional areas responsible for SdH quantum oscillations were determined with SKEAF package\cite{rourkeNumericalExtractionHaas2012}.

\section{RESULTS AND DISCUSSIONS}

\begin{figure*}[!htbpb]
	\includegraphics[width= 12cm]{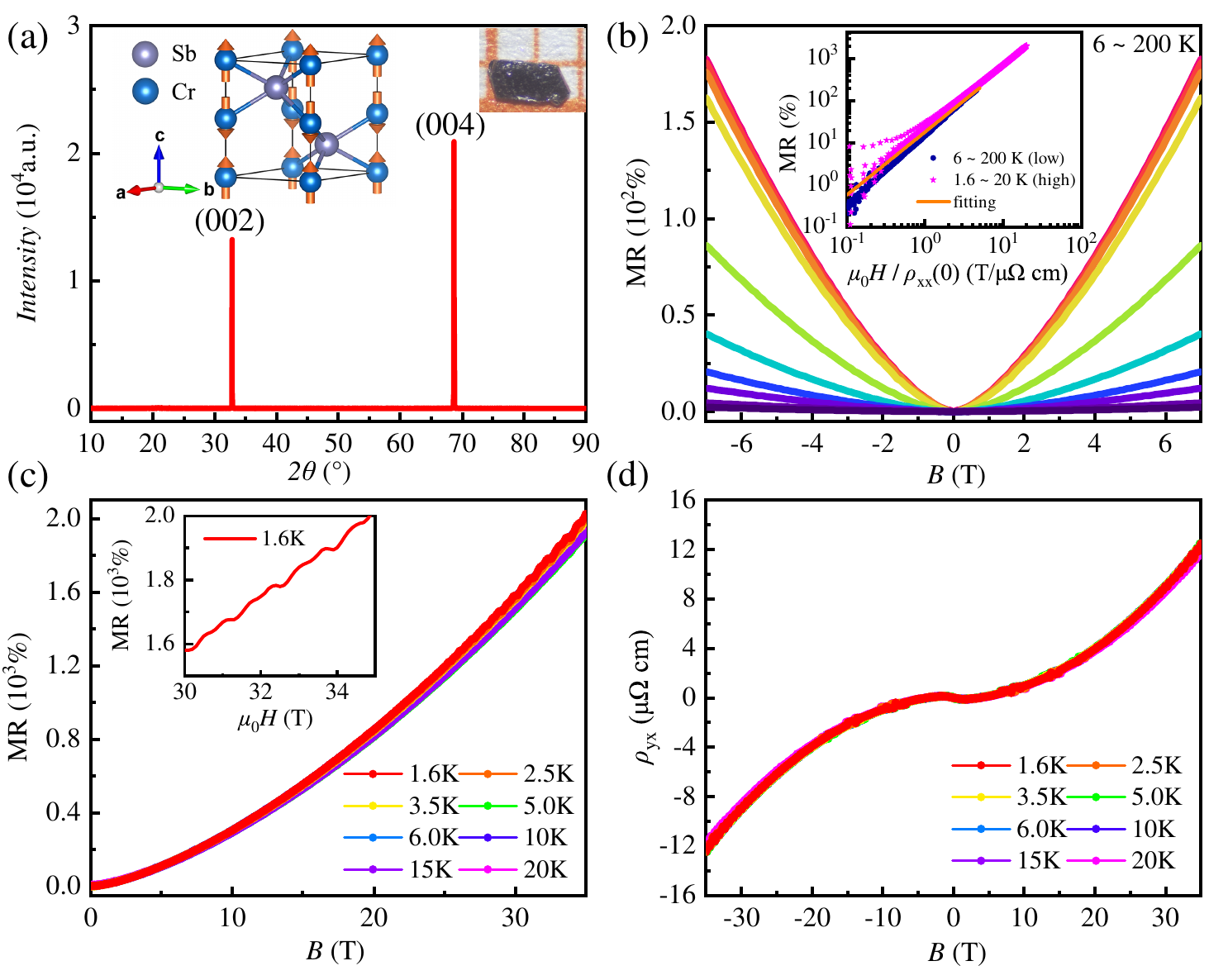}
	\caption{\label{FIG. 1}(a) Single-crystal XRD pattern. The inset on the upper left: a schematic diagram of the magnetic moment configuration. The inset on the upper right: an image of the single crystal. (b) Field-dependent MR curves measured at various temperatures in a magnetic field range of 0-7 T. The inset: the Kohler’s law fitting results for both low-field and high-field regions. (c) MR curves as a function of magnetic field at different temperatures under high-field measurements. The inset: an enlarged view of the MR curve at 1.6 K. (d)Field-dependent $\rho_{yx}$ curves at different temperatures under high magnetic fields.}
\end{figure*}

Figure 1(a) shows the single-crystal x-ray diffraction (XRD) pattern, in which only (00l) peaks with a narrow full width at half maximum ($\sim$0.05$^{\circ}$) are observed, confirming the high crystalline quality of the sample. The inset illustrates the hexagonal NiAs-type structure of CrSb, in which magnetic moments align ferromagnetically within the $ab$ plane and antiferromagnetically between layers along the $c$-axis, consistent with an A-type antiferromagnetic ordering\cite{PhysRev.85.365,10.1063/5.0158271,PhysRev.129.2008}. Figure 1(b) presents the magnetoresistance (MR) as a function of the magnetic field at various temperatures. MR is defined as MR = [$\rho(\mu_{0}H)$ - $\rho(0)$]/${\rho(0)}\times100\%$, where $\rho(\mu_{0}H)$ and $\rho(0)$ are the resistivities measured with and without an applied magnetic field, respectively. Non-saturating MR behavior is observed up to 7 T, reaching 183\% at 6 K, which is approximately 3.5 times larger than the MR of our first-generation samples\cite{PhysRevB.111.144402}. Subsequently, we performed magnetotransport measurements in pulsed fields up to 35 T. As shown in Figure. 1(c), the MR increases to more than 2000\% at 1.6 K and 35 T, with no sign of saturation, which is a typical behavior for
many trivial or nontrivial semimetals with XMR\cite{PhysRevB.97.245101,PhysRevB.94.235154,PhysRevB.102.165133,SciChinaPhysMechAstron.59.037411,ChinPhysB.33.037301,PhysRevB.102.115145,PhysRevB.104.115104}. The magnetoresistance behavior can be described by Kohler’s rule\cite{Pippard1989,https://doi.org/10.1002/andp.19384240124}:
\begin{equation}
	\mathrm{MR} = \frac{\triangle\rho_{xx}(T, H)}{\rho_{0}(T)} = \alpha(\frac{B}{\rho_{0}(T)})^m
\end{equation}     
As shown in the inset of Figure. 2(b), the magnetoresistance curves from two independent measurements (high-field and low-field regimes) collapse onto a single straight line after normalization. The fitting yields m = 1.48 and $\alpha$ = 0.19 ($\mu\Omega$ cm/T)$^{1.48}$. Figure 1(d) displays the field-dependent Hall resistivity, showing nonlinear behavior at various temperatures. Combined with our initial sample measurements and analysis\cite{PhysRevB.111.144402}, this suggests multiband charge carrier effects. The high-field magnetoresistance exhibits periodic oscillations, with prominent Shubnikov-de Haas (SdH) quantum oscillations visible in the inset of Figure. 1(c), suggesting further investigation.

After subtracting the smooth background of the high-field magnetoresistance, clear SdH oscillations were obtained. The oscillatory component $\triangle$$\rho_{xx}$ as a function of the reciprocal magnetic field (1/$B$) at various temperatures is plotted in Figure. 2(a). To further investigate the influence of high magnetic fields on the Fermi surface of CrSb, we performed fast Fourier transform (FFT) analyses of the SdH oscillations at 1.6 K over multiple field windows, as shown in the inset of Figure. 2(b)\cite{PhysRevLett.114.176601,PhysRevB.92.125152}. The FFT spectra extracted from four different field ranges (26 - 32 T, 30 - 35 T, 28 - 35 T, 24 - 35 T). In the broadest field window (24 - 35 T), multiple frequency components can be clearly resolved, including a fundamental frequency $F_{1}$, its higher harmonics (2$F_{1}$, 3$F_{1}$), and two additional prominent peaks $F_{2}$ and $F_{3}$, located at 76.35 T, 839.9 T, and 1603.4 T, respectively. The corresponding Fermi surface cross-sectional areas are extracted using the Onsager relation F = $S_{F}$($\hbar$/2$\pi$e), yielding $S_{F_{1}}$ = 0.00729 \AA$^{-2}$, $S_{F_{2}}$ = 0.08013 \AA$^{-2}$, and $S_{F_{3}}$ = 0.15314 \AA$^{-2}$. As the magnetic field increases and the field window narrows, the amplitude of $F_{1}$ gradually diminishes and nearly vanishes within the 30 - 35 T range, while $F_{2}$ and $F_{3}$ become progressively more pronounced. This field-dependent behavior indicates a strong sensitivity of the SdH oscillations in CrSb to the applied magnetic field and suggests the possible reconstruction of the Fermi surface under high fields, potentially due to the Zeeman effect. To further elucidate the Fermi surface properties, we also carried out temperature-dependent FFT analyses, as shown in Figure. 2(b).

Subsequently, the quantum oscillations were well described using the Lifshitz–Kosevich (LK) formula\cite{NatCommun.7.12516,PhysRevB.103.165128,PhysRevB.109.165155,SciRep.13.22776,PhysRevB.98.115145},
\begin{equation}
	\triangle\rho \propto R_{T}R_{D}R_{S}\ \mathrm{cos}[2\pi(\frac{F}{B} + \gamma - \delta)]
\end{equation}   
Where $R_{T}$ = $\frac{2\pi^{2}k_{B}m_{0}\mu T}{\hbar eB}$/sinh$\frac{2\pi^{2}k_{B}m_{0}\mu T}{\hbar eB}$ is the thermal damping factor accounting for the suppression of oscillation amplitude with increasing temperature. Here, $\mu$ = $\frac{m^{*}}{m_{0}}$, with $m^{*}$ being the effective mass, $m_{0}$ the free electron mass, $K_{B}$ the Boltzmann constant, $\hbar$ the reduced Planck constant, and $B$ the average magnetic field over the FFT window\cite{npjQuantMater.5.10,NatCommun.9.3249}. The Dingle damping factor $R_{D}$ = $e^{-D}$ reflects the decay due to impurity scattering, where D = $\frac{2\pi^{2}k_{B}T_{D}m^{*}}{\hbar eB}$, and the Dingle temperature is defined as $T_{D}$ = $\frac{\hbar}{2\pi k_{B}\tau_{q}}$, with $\tau_{q}$ the quantum scattering time. The spin damping factor is given by $R_{S}$ = cos($\pi$g$m^{*}$/2$m_{0}$), where g is the Landé $g$-factor, typically taken as 2\cite{PhysRevB.102.115145}. The phase shift $\gamma-\delta$ includes a geometric component $\gamma$ = 1/2 - $\Phi/2\pi$, where $\Phi$ is the Berry phase, and a dimensional correction $\delta$, which takes the values of 0 in two-dimensional systems, and $\pm\frac{1}{8}$ in three-dimensional cases\cite{doi:10.1126/science.1242247}. To extract the effective masses of the charge carriers associated with each frequency, the temperature dependence of the FFT amplitudes was fitted using the thermal damping term $R_{T}$, as shown in Figure. 2(c). The effective masses were determined to be $m^{*}_{1}$ = 0.867$m_{0}$, $m^{*}_{2}$ = 1.5$m_{0}$, and $m^{*}_{3}$ = 0.945$m_{0}$, corresponding to the three main frequencies observed. Using these values, the Fermi velocity can be estimated via $\upsilon_{F}$ = $\hbar k_{F}/m^{*}$, resulting in 2.034 $\times$ $10^{4}$ m/s, 1.21 $\times$ $10^{5}$ m/s, and 2.69 $\times$ $10^{5}$ m/s, respectively. The Fermi wave vector is calculated from $k_{F}$ = $\sqrt{2eF/\hbar}$, and the Fermi energy is evaluated as $E_{F}$ = $\upsilon^{2}_{F}m^{*}$, giving $E_{F}$ = 0.61 meV, 124.8 meV, and 388.6 meV, respectively. We then fitted the oscillatory curves at 1.6 K using the LK formula with the extracted parameters, as shown in Figure. 2(d)\cite{PhysRevB.109.165155}. For the frequency $F_{1}$, a phase shift of $\gamma-\delta$ = 0.1416 was obtained, yielding Berry phases $\Phi$ = 0.4578$\pi$ for $\delta$ = +$\frac{1}{8}$ and $\Phi = 0.9578\pi$  for $\delta$ = -$\frac{1}{8}$ . Similarly, for $F_{2}$, the extracted phase shift was $\gamma-\delta$ = 0.1047, giving $\Phi$ = 0.5406$\pi(\delta$ = +$\frac{1}{8})$ and $\Phi$ = 1.0406$\pi(\delta$ = -$\frac{1}{8})$. For $F_{3}$, the corresponding phase shift  $\gamma-\delta$ = 0.0813 leads to $\Phi$ = 0.5874$\pi$ and $\Phi$ = 1.0874$\pi$, respectively. The Dingle temperatures $T_{D}$ extracted from fitting are 48.397 K, 6.165 K, and 11.876 K, from which the quantum scattering times $\tau$ are calculated to be $2.5 \times 10^{-14}$ s, $1.97 \times 10^{-13}$ s, and $1.02 \times 10^{-13}$ s, respectively. These yield corresponding quantum mobilities $\mu_{q} = e\tau_{q}/m^{*}$ of 50.65, 230.7, and 189.4 cm$^{2}$/V$\cdot$s. The mean free paths $\ell$ = $\upsilon_{F}\tau_{q}$ are estimated to be 0.51 nm, 23.84 nm, and 27.4 nm, respectively. A complete summary of these parameters is provided in Table I.

\begin{table}[htbp!]
	\centering
	\caption{The parameters extracted from SdH oscillation fitting of CrSb, including the oscillation frequencies $F_{1}$, $F_{2}$, and $F_{3}$, as well as their corresponding effective mass ratio $m^{*}/m_{0}$, Fermi wave vector $K_{F}$, Fermi velocity $\upsilon_{F}$, Fermi energy $E_{F}$, Dingle temperature $T_{D}$, quantum scattering time $\tau_{q}$, quantum mobility $\mu_{q}$, and mean free path $\ell$.}
	\setlength{\tabcolsep}{6.7mm}{
		\begin{tabular}{@{}l  c c c@{}}
			\hline
			\hline
			Parameters & $F_{1}$ & $F_{2}$ & $F_{3}$ \\
			\hline
			
			Frequency (T) & 76.35 & 839.9 & 1603.4 \\
			
			$m^{*}/m_{0}$ & 0.867 & 1.5 & 0.945 \\
			
			$K_{F}$ (\AA$^{-1}$) & 0.015 & 0.157 & 0.221 \\
			
			$\upsilon_{F}$ ($10^{5}$m/s) & 0.2 & 1.21 & 2.69 \\
			
			$E_{F}$ (meV) & 0.61 & 124.8 & 388.6 \\
			
			$T_{D}$ (K) & 48.397 & 6.165 & 11.876 \\
			
			$\tau_{q}$ ($10^{-13}$s) & 0.25 & 1.97 & 1.02 \\
			
			$\mu_{q}$ (cm$^{2}/$/V$\cdot$s) & 50.56 & 230.7 & 189.4 \\
			
			$\ell$ (nm) & 0.51 & 23.84 & 27.4 \\
			\hline
			\hline
	\end{tabular}}
\end{table}

To further investigate the topological characteristics of CrSb, in addition to the Berry phases extracted from the LK analysis, we performed an independent verification via Landau level (LL) index fan diagrams\cite{doi:10.1126/science.1242247,doi:10.1126/science.1189792,PhysRevB.98.195136}. To suppress interference from multiple frequency components, the SdH oscillation signals were processed using band-pass filtering. We selectively retained narrow spectral windows centered on the individual frequencies $F_{1}$, $F_{2}$, and $F_{3}$, thereby isolating single-frequency oscillations for each case, as presented in Figure. 2(d). We then assigned Landau level indices to the filtered oscillation traces by identifying the positions of local extrema (either peaks or valleys) in the SdH oscillations. This was done under the Lifshitz-Onsager(LO) formula, expressed as $S_{F}(\hbar/eB) = 2\pi(n + \gamma - \delta)$\cite{PhysRevB.109.165155}, where $n$ is the Landau level index. According to convention, adjacent maxima (or minima) in the oscillatory component correspond to consecutive integer indices, while a maximum and its neighboring minimum differ by 0.5. Based on this assignment, a plot of Landau index $n$ versus the inverse magnetic field $1/B$ was constructed, as shown in Figure. 2(f), providing an independent means to evaluate the Berry phase. By extrapolating the Landau index plots to the zero inverse field limit, we determined the intercepts $n_{0}$ for each frequency component, For the oscillation frequency $F_{1}$, the extracted intercept was $n_{0}$ = -0.2280, corresponding to Berry phases of $\Phi = 0.2940\pi$ when $\delta$ = +$\frac{1}{8}$, and $\Phi = 0.7940\pi$ when $\delta$ = -$\frac{1}{8}$. Similarly, for $F_{2}$, an intercept of $n_{0}$ = -0.5193 yielded Berry phases of -0.2885$\pi$ and 0.2115$\pi$ for$\delta$ = +$\frac{1}{8}$ and$\delta$ = -$\frac{1}{8}$, respectively. The third frequency, $F_{3}$, gave an intercept of $n_{0}$ = -0.1667, which corresponds to $\Phi = 0.4167\pi$ for $\delta$ = +$\frac{1}{8}$ and $\Phi = 0.9167\pi$ for $\delta$ = -$\frac{1}{8}$. Notably, the Berry phases corresponding to the frequency components $F_{1}$ and $F_{3}$ exhibit good consistency between the LL index fitting and the LK fitting methods. In particular, under the LL fitting with a phase offset of $\delta$ = -$\frac{1}{8}$, the extracted Berry phases for both components approach $\pi$, the comparative data are shown in Table 2. In conjunction with previous reports, this feature suggests that the Fermi surface pockets associated with $F_{1}$ and $F_{3}$ may originate from Weyl-type band structures with linear energy dispersion. This result further supports the possible existence of topologically nontrivial electronic states in CrSb.

\begin{figure*}
	\includegraphics[width= 17cm]{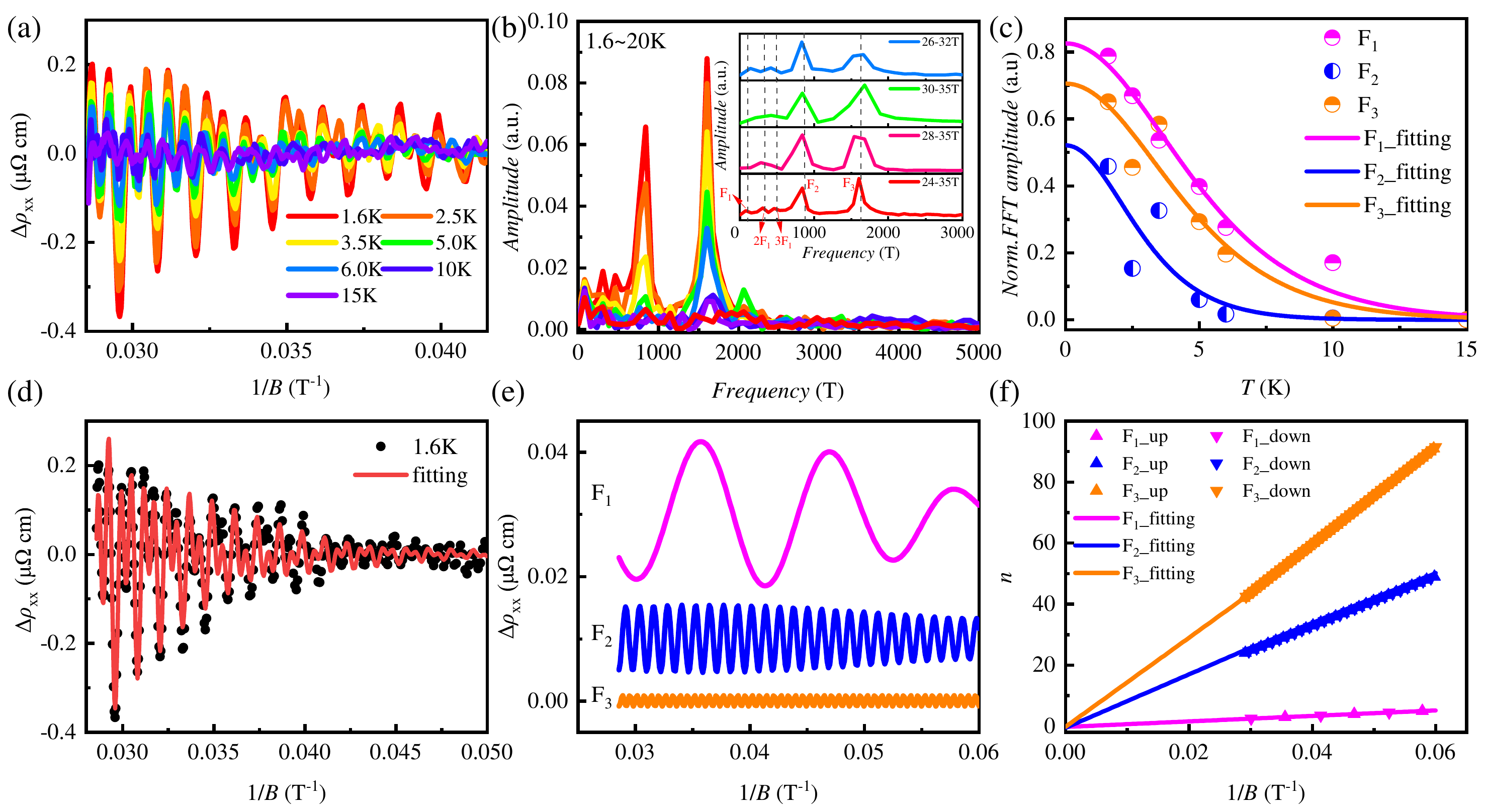}% Here is how to import EPS art
	\caption{\label{FIG. 2}(a) SdH oscillatory components at various temperatures plotted as a function of the inverse magnetic field (1/$B$) after background subtraction. (b) FFT spectra of the SdH oscillations measured at different temperatures, obtained within the magnetic field range of 24-35 T. The inset: the FFT spectra at 1.6 K over four different field windows, from which the fundamental frequency $F_{1}$ = 76.35 T and its higher harmonics, as well as frequencies  $F_{2}$ = 839.9 T and $F_{3}$ = 1603.4 T are identified. (c) Temperature dependence of the normalized FFT amplitudes for the three frequencies. Solid curves represent fits using the Lifshitz-Kosevich (LK) formula. (d) SdH oscillatory component of $\triangle\rho_{xx}$ measured at 1.6 K, along with the fitting curve based on a multiband LK formula. (e) Band-pass filtered oscillations corresponding to the three fundamental frequencies. (f) Landau-level index fan diagram for the three frequencies, showing the linear relation between the Landau index $n$ and inverse magnetic field 1/$B$.}  
\end{figure*}

\begin{table}[htbp!]
	\centering
	\caption{The comparison of the Landau-Kosevich (LK) and Lifshitz-Kosevich (LL) fitting results for Shubnikov-de Haas (SdH) oscillations.}
	\setlength{\tabcolsep}{4.5mm}{
		\begin{tabular}{@{}l  c c c@{}}
			\hline
			\hline
			Parameters (LK) & $F_{1}$ & $F_{2}$ & $F_{3}$ \\
			\hline
			
			$\gamma$ (1/8) & 0.2711 & 0.2297 & 0.2063 \\
			
			$\Phi$ (1/8) & 0.4578$\pi$ & 0.5406$\pi$ & 0.5874$\pi$ \\
			
			$\gamma$ (-1/8) & 0.0211 & -0.0203 & -0.0437 \\
			
			$\Phi$ (-1/8) & 0.9578$\pi$ & 1.0406$\pi$ & 1.0874$\pi$ \\
			\hline
			\hline
			Parameters (LL) & $F_{1}$ & $F_{2}$ & $F_{3}$ \\
			\hline
			
			$n_{0}$ & -0.2280 & -0.5193 & -0.1667 \\
			
			$\Phi$ (1/8) & 0.2940$\pi$ & -0.2885$\pi$ & 0.4167$\pi$ \\
			
			$\Phi$ (-1/8) & 0.7940$\pi$ & 0.2115$\pi$ & 0.9167$\pi$ \\
			\hline
			\hline
	\end{tabular}}
\end{table}

\begin{figure*}
	\includegraphics[width= 14cm]{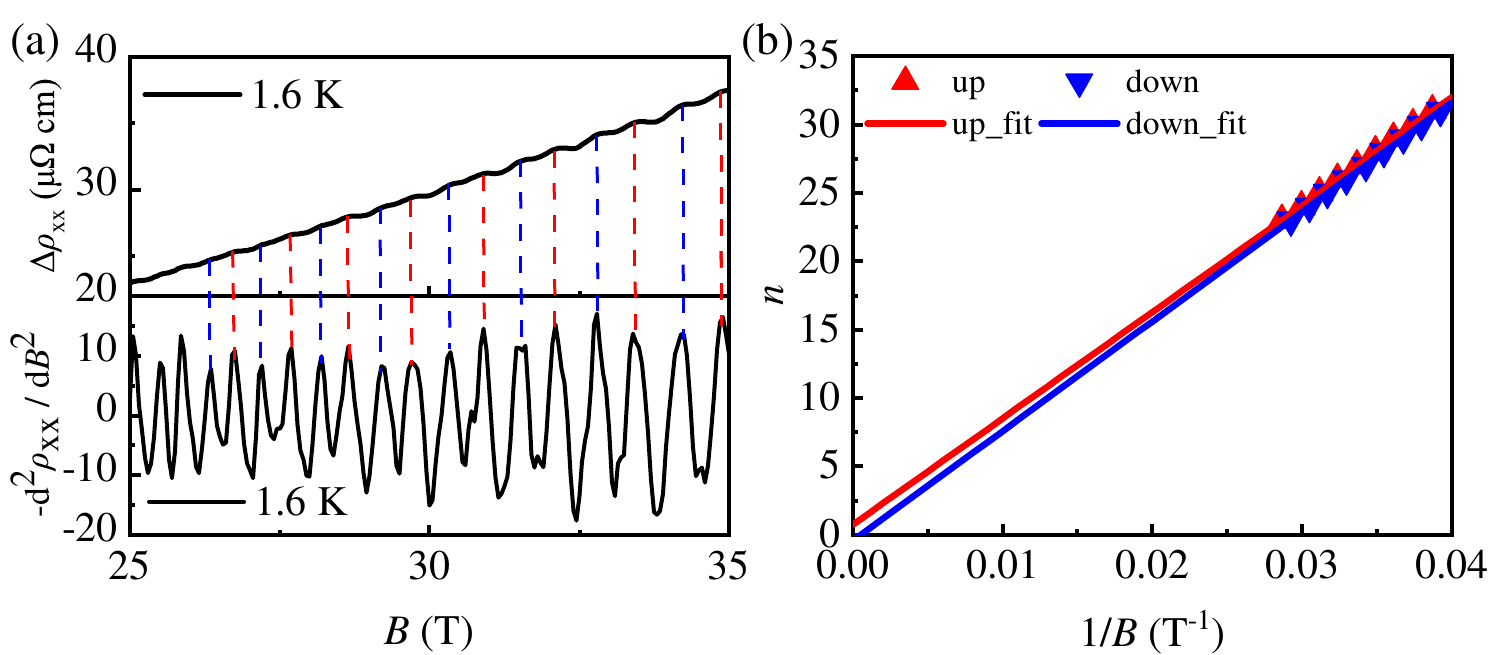}% Here is how to import EPS art
	\caption{\label{FIG. 3} (a) The $\triangle\rho_{xx}$ curve as a function of magnetic field and the corresponding -$d^{2}\rho_{xx}/dB^{2}$ curve at 1.6 K. (b) Landau-level fan diagram exhibiting Zeeman splitting.}
\end{figure*}

In addition to the aforementioned quantum oscillation behavior, signatures of the Zeeman effect were also observed. As shown in Figure. 3(a), the upper panel shows the variation of $\triangle\rho_{xx}$ as a function of magnetic field, while the lower panel presents the corresponding second derivative curve, -$d^{2}\rho_{xx}/dB^{2}$. In the $\triangle\rho_{xx}$ oscillation signal, two distinct peaks are observed within each period. To more clearly resolve the Zeeman splitting features in the SdH oscillations, the second derivative of the longitudinal resistivity with respect to the magnetic field, $\triangle\rho_{xx}(B)$, is taken and its negative value is plotted. As discussed by Su $et\ al$ for  LaAlSi\cite{PhysRevB.103.165128}, this method enhances the contrast of local maxima and minima, allowing the main oscillation peaks to be distinguished from the spin-split sub-peaks, thereby enabling a more accurate extraction of the Landau level splitting behavior. Notably, the oscillation peaks marked by dashed lines in Fig. 3(a) exhibit a clear double-peak splitting structure at high magnetic fields, while at lower fields (below $\sim$28 T), the splitting width gradually increases. Moreover, as shown in Figs. 3(a) and 3(d), the splitting gradually diminishes with increasing magnetic field. This behavior indicates that the oscillation signal under strong magnetic fields is influenced by spin splitting, exhibiting the characteristic Zeeman splitting feature. Therefore, the spin damping factor $R_{x}$ = cos$(g\pi m^{*}/2m_{0})$ is used to describe spin degeneracy, and the previously mentioned LK formula needs to be modified accordingly as follows\cite{PhysRevB.109.165155,PhysRevB.103.165128,NatCommun.7.12516}:
\begin{equation}
	\triangle\rho \propto R_{T}R_{D}\ \mathrm{cos}[2\pi(\frac{F}{B} + \gamma - \delta + \frac{\varphi}{2}) 
	+ \frac{F}{B} + \gamma - \delta - \frac{\varphi}{2})]
\end{equation}   
where $\varphi = \pi gm^{*}/m_{0}$ represents the phase difference between the spin-split components caused by the Zeeman effect. To further quantify the spin splitting behavior, we constructed Landau fan diagrams based on the split peaks, as shown in Figure. 3(b). The upper and lower branches of the split peaks were assigned distinct Landau level indices (represented by red and blue symbols, respectively), and linear fits were performed separately. By comparing the slopes and intercepts of these fits, we extracted the corresponding spin-related phase offsets. The effective Landé g factors were determined to be $g$ = 2.68 for frequency $F_{1}$, $g$ = 1.55 for $F_{2}$, and $g$ = 2.46 for $F_{3}$.

% ------------------------------------------------------------

\begin{figure*}
	\includegraphics[width= 13cm]{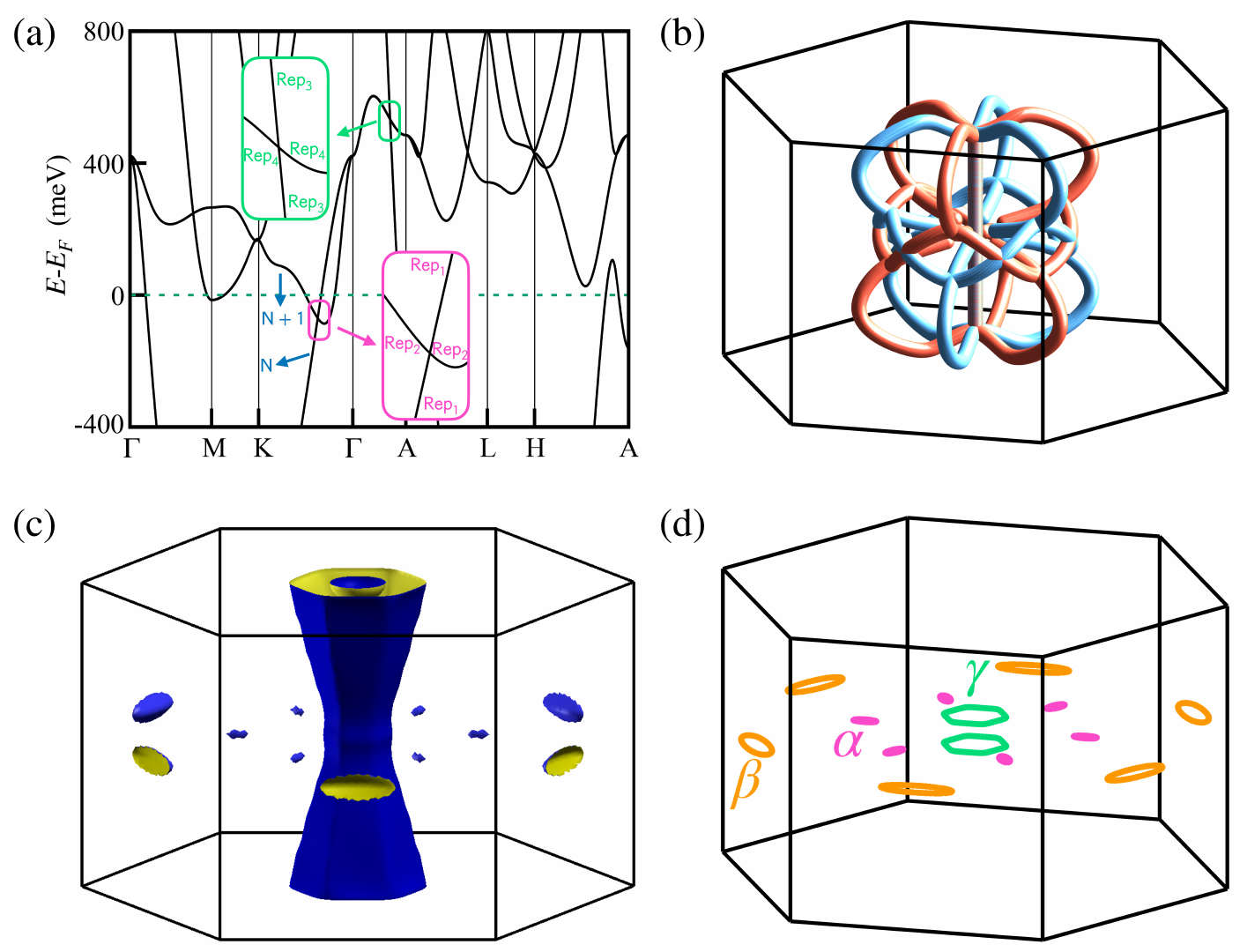}
	\caption{\label{FIG. 4} (a) Band structure of CrSb without spin-orbit coupling (SOC) along the high-symmetry path $\Gamma$-M-K-$\Gamma$-A-L-H-A. The red and blue frames highlight two sets of symmetry-protected band crossings. (b) Distribution of the nodal line in the first Brillouin zone without SOC; red (blue) segments denote spin-up (spin-down) points\cite{luSignatureTopologicalSurface2025}. (c) Selected fermi-surface including SOC, showing only the pockets of those bands that contribute to quantum-oscillation signals. (d) Cross-section of the SOC Fermi surface in (c), highlighting the distinct extremal orbits $\alpha$, $\beta$ and $\gamma$ that are expected to dominate quantum-oscillation signals.}
\end{figure*}

Figure \ref{FIG. 4}(a) shows the band structure of CrSb calculated without SOC. Along the $K - \Gamma$ high-symmetry line (pink rectangle) a symmetry-protected intersection arises because the two branches belong to different irreducible representations, Along $\Gamma - A$, the pair of bands, marked by the green-colored $\mathrm{Rep}_4$, remains degenerate throughout the segment, protected by symmetry, as obtained from TABLE \ref{table 3}, and the representations are calculated based on spin space group theory\cite{jiangEnumerationSpinSpaceGroups2024, songConstructionsApplicationsIrreducible2025}. These degeneracies persist until the corresponding symmetry is broken. Figure \ref{FIG. 4}(b) maps the nodal line formed by the crossing of bands with occupation numbers $N$ and $N + 1$ (as indicated in Figure \ref{FIG. 4}(a)) within the first Brillouin zone, without SOC. Red (blue) segments represent spin-up (spin-down) points along the line\cite{luSignatureTopologicalSurface2025}. Introducing SOC gaps the nodal lines and reconstructs the Fermi surface. Figure \ref{FIG. 4}(c) plots only those pockets whose extremal cross-sectional areas are expected to give quantum-oscillation signals. Figure \ref{FIG. 4}(d) presents cross-section of the SOC Fermi surface, revealing three principal extremal orbits, labelled $\alpha$, $\beta$ and $\gamma$. The symmetry-protected crossings highlighted in Figure. \ref{FIG. 4}(a) pass through the $\alpha'$ and $\gamma'$ contours in the weak SOC limit. As shown in Fig. \ref{Appendix 1} of the appendix, these pockets evolve into the $\alpha$ and $\gamma$ pockets in Figure. \ref{FIG. 4}(d) upon the inclusion of SOC. The Berry phase, originally generated by the crossings, is retained in the Fermi surface sections of the $\alpha$ and $\gamma$ pockets in the SOC case.

\begin{table}[htbp!]
	\centering
	\caption{\label{table 3}Symmetry operators associated with the four band-crossing states highlighted in Figure. \ref{FIG. 4}(a). Only those symmetry operations that distinguish the two irreducible representations in each pair are listed. Each entry in the first column is written in the compact form $\{\,U_i \,\|\, R_i \mid \tau_i\}$, where $U_i$ denotes the spin-rotation part, $R_i$ the corresponding real-space rotation/mirror acting on the lattice, and $\tau_i$ the accompanying translation vector. Column headers $\mathrm{Rep}_{1}(2)$, $\mathrm{Rep}_{2}(2)$, $\mathrm{Rep}_{3}(2)$, and $\mathrm{Rep}_{4}(4)$ label the four bands; the number in parentheses gives the degeneracy (dimension) of that band at the indicated $k$-path segment. The upper block lists results for the $K\!-\!\Gamma$ path, while the lower block corresponds to the $\Gamma\!-\!A$ path.}
	\setlength{\tabcolsep}{7mm}{
		\begin{tabular}{@{}l  c c@{}}
			\hline
			\hline
			Representations ($K-\Gamma$) & $\mathrm{Rep}_{1}(2)$ & $\mathrm{Rep}_{2}(2)$ \\
			\hline
			
			$\{E||C_{2(110)}|0,0,0\}$ & -2 & 2 \\
			
			\hline
			\hline
			Representations ($\Gamma-A$) & $\mathrm{Rep}_{3}(2)$ & $\mathrm{Rep}_{4}(4)$ \\
			\hline
			
			$\{E||E|0,0,0\}$ & 2 & 4 \\
			
			$\{E||C_{3z}|0,0,0\}$ & 2 & -2 \\
			
			$\{E||C^2_{3z}|0,0,0\}$ & 2 & -2 \\
			
			$\{E||m_{110}|0,0,0\}$ & 2 & 0 \\
			
			$\{E||m_x|0,0,0\}$ & 2 & 0 \\
			
			$\{E||m_y|0,0,0\}$ & 2 & 0 \\
			
			\hline
			\hline  
	\end{tabular}}
\end{table}

% -----------------------------------------------------------

~\\
\section{CONCLUDING REMARKS}

In summary, we have successfully grown high-quality CrSb single crystals and systematically investigated their electrical transport properties. The magnetoresistance shows no sign of saturation up to 35 T and follows a $B^{1.48}$ power-law dependence. The nonlinear behavior of the Hall resistivity suggests multiband charge transport. Most remarkably, we observe pronounced SdH oscillations in high magnetic fields. Combined first-principles calculations and experimental analyses confirm the existence of nontrivial Berry curvature in CrSb, indicating the participation of topologically protected electronic states in the quantum oscillations. Furthermore, distinct Zeeman splitting is observed in the SdH oscillations at 1.6 K, manifested as spin splitting of Landau levels. These findings not only deepen our understanding of the quantum transport properties in CrSb but also establish this material as an intriguing candidate for studying topological phenomena in altermagnet.

~\\
\begin{acknowledgments}
	
This research is supported by the National Key R$\&$D program of China (Grant No. 2022YFA1403202, No. 2023YFA1607400, No. 2024YFA1408400), and the National Natural Science Foundation of China (Grant No. 12074335, No. 52471020, No. 12274436, No. 12188101). We thank the WM5 (https://cstr.cn/31125.02.SHMFF.WM5) at the Steady High Magnetic Field Facility, CAS (https://cstr.cn/31125.02.SHMFF), for providing technical
support and assistance in data collection and analysis.

\end{acknowledgments}

%\renewcommand{\thefigure}{S\arabic{figure}}

\nocite{*}

\bibliography{Reference}% Produces the bibliography via BibTeX.
\cleardoublepage
\appendix
\onecolumngrid

\section{Fermi-surface evolution from weak to full spin–orbit coupling}

\begin{figure*}[h]
	\includegraphics[width= 14cm]{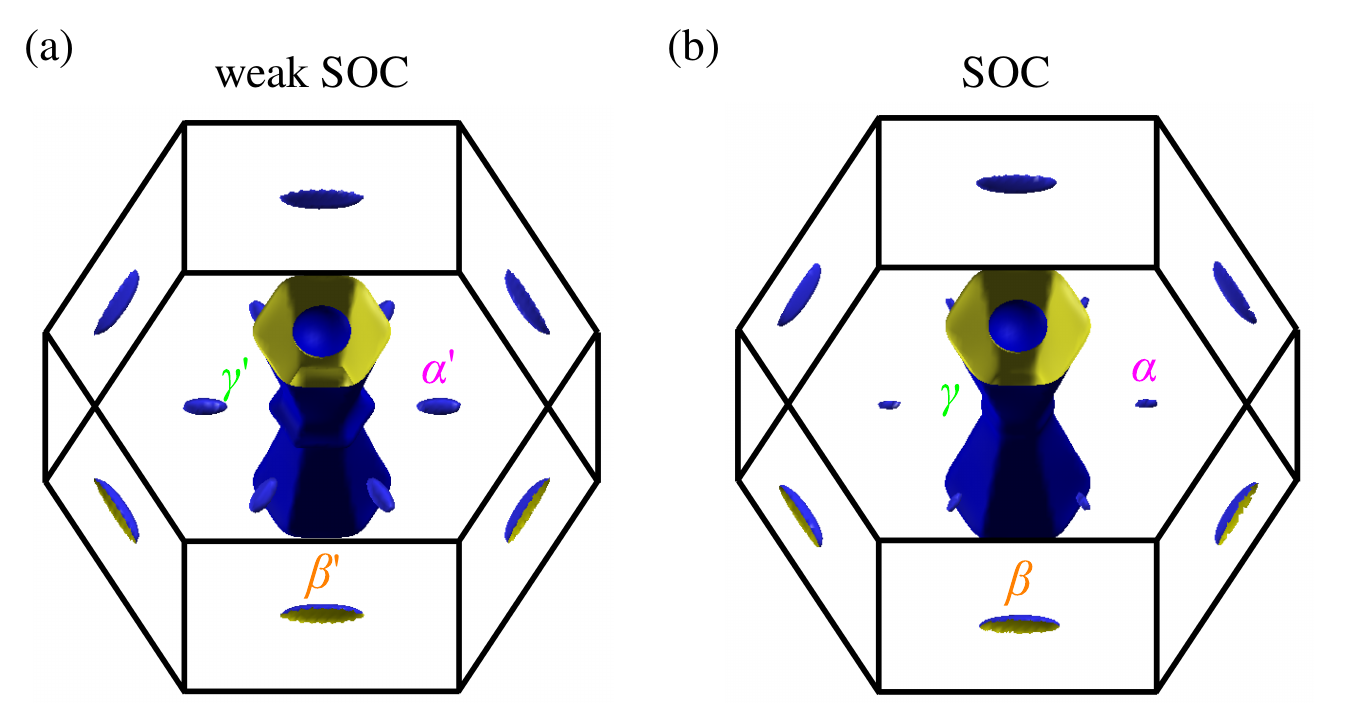}
	\caption{\label{Appendix 1} Evolution of the selected fermi surface from weak SOC to SOC. (a) Selected fermi surface in the weak SOC limit, showing the $\alpha'$, $\beta'$, and $\gamma'$ pockets. These pockets are located at different positions and sizes in the Brillouin zone. (b) Upon including spin-orbit coupling (SOC), the $\alpha'$, $\beta'$, and $\gamma'$ pockets transform into the $\alpha$, $\beta$, and $\gamma$ pockets, respectively. }
\end{figure*}

~\\
\noindent\textbf{Author contributions}
~\\
\noindent  J. D., X. P., Y. W. contributed equally to this work. J. D., Z. J., Q. W., and M. F. conceived of and designed the study. X. P., Y. Sun performed the crystal growth; X. P., C. W., T. Z., L. L., H. W., J. Y., and B. C. performed magnetization and transport measurements; X. P. and C. X. performed transport measurements on the water-cooled magnet, Y. W., S. Z. and Q. W. carried out the theoretical calculation; X. P., J. D., Q. W.and M. F. performed fundamental data analysis; J. D., X. P., Y. W., Q. W. and M. F. wrote the manuscript based on discussion with all the authors.

\end{document}